\documentclass[showpacs,twocolumn,preprintnumbers,amsmath,amssymb]{revtex4-1}
\usepackage[english]{babel}
\usepackage{ucs}
\usepackage[utf8x]{inputenc}
\usepackage[pdftex]{graphicx}

\begin{document}

\title{Structural manipulation of the graphene/metal-interface with
Ar$^+$ irradiation}

\author{E. H. {\AA}hlgren$^1$,
S. K. H\"am\"al\"ainen$^2$,
O. Lehtinen$^3$,
P. Liljeroth$^2$, and
J. Kotakoski$^{1,4}$}
\email{jani.kotakoski@iki.fi}

\affiliation{$^1$ Department of Physics, University of Helsinki, P.O. Box 43,
FI-00014, Finland\\
$^2$ Department of Applied Physics, Aalto University School of Science, P.O.
Box 15100, FI-00076, Finland\\
$^3$ Facility for Electron Microscopy, Group of Electron Microscopy of
Materials Science, University of Ulm, 89081 Ulm, Germany\\
$^4$ Faculty of Physics, University of Vienna, Boltzmanngasse 5,
1090 Vienna, Austria}


\date{\today}

\begin{abstract}

Controlled defect creation is a prerequisite for the detailed study of disorder
effects in materials. Here, we irradiate a graphene/Ir(111)-interface with
low-energy Ar$^+$ to study the induced structural changes.  Combining computer
simulations and scanning-probe microscopy, we show that the resulting disorder
manifests mainly in the forms of intercalated metal adatoms and vacancy-type
defects in graphene. One prominent feature at higher irradiation energies (from
1~keV up) is the formation of line-like depressions, which consist of
sequential graphene defects created by the ion channeling within the
interface---much like a stone skipping on water.  Lower energies result in
simpler defects, down to 100~eV where more than one defect in every three is a
graphene single vacancy.

\end{abstract}

\maketitle

% INTRO

\section{Introduction}

Measurements of the properties of graphene, especially the electronic transport
characteristics~\cite{castro_neto_electronic_2009}, have yielded results which
deviate from the theory~\cite{geim_rise_2007}. These discrepancies are likely
to have partially risen from disorder, introduced either already in growth or
later during sample preparation. A detailed correlation between the
atomic-level disorder and the corresponding change in the properties requires a
controlled way of creating defects. On the other hand, defects can also be
beneficial for nanoengineering low-dimensional
structures~\cite{krasheninnikov_engineering_2007}.

One way for manipulating structures is to employ beams of energetic particles,
for example ions.  While ion irradiation effects in conventional materials have
been studied for decades, we are only starting to fully understand the
irradiation-response of low-dimensional structures. In the case of suspended
graphene, some of the authors of the present work have recently conducted
atomic-level theoretical studies of irradiation effects for various
ions~\cite{lehtinen_effects_2010,lehtinen_cutting_2011,ahlgren_ion_2012}. For
supported graphene, experimental studies have been carried out for irradiation
with 140 eV Ar$^+$ on Pt(111)~\cite{ugeda_point_2011} and
SiC(0001)~\cite{ugeda_electronic_2012} surfaces, 30~keV and 100~keV Ar$^+$ on
SiO2 substrate~\cite{tapaszto_tuning_2008,kalbac_ion-irradiation-induced_2013},
and 5~keV Xe$^+$ on Ir(111) surface (at a grazing
angle)~\cite{standop_ion_2013}. Ion irradiation has already been demonstrated
to tune the electronic and magnetic properties of
graphene~\cite{tapaszto_tuning_2008,chen_defect_2009,nair_spin-half_2012,nair_dual_2013},
and to create single~\cite{ugeda_point_2011} and double
vacancies~\cite{ugeda_electronic_2012}, although no direct correlation between
the properties and defect structures has been carried out. Overall, as
far as we know, no atomic-level analysis of irradiation-induced structural
changes at varying irradiation energies has been reported for supported
graphene.

In this study, we investigate Ar$^+$ irradiation effects on graphene on a
weakly interacting metal substrate, Ir(111), by combining atomistic simulations
and scanning tunneling microscopy (STM).  Good qualitative agreement between
the methods (over all experimentally studied irradiation energies: 0.1~keV,
0.3~keV and 1~keV) allows us to assign the experimentally observed disorder to
intercalated adatoms and disordered graphene areas. According to simulations,
as compared to suspended graphene, the substrate leads to more complex defect
structures at all studied energies (up to 30~keV). Nevertheless, at the lowest
studied energy above the damaging threshold (100~eV), more than every third
created defect is a single vacancy in graphene.  Below 1~keV the substrate
somewhat protects graphene, whereas above 1~keV backscattered metal atoms
increase the damage. At similar energies, ion channeling below the graphene
sheet starts to occur, leading to line-like features of small vacancy-type
graphene defects.  Such large variations in the introduced disorder as a
function of the ion energy, already at the relatively small energy range of
100~eV---1~keV, show a great promise for ion irradiation-mediated manipulation
of graphene/metal-interfaces.

% EXPERIMENTS

\section{Methods and results}

We used a Createc LT-STM system for the Ar+ irradiation experiments (base
pressure around $1\times 10^{−10}$~mbar). Following the procedure outlined in
Ref.~\cite{coraux_growth_2009}, graphene growth was initiated by absorbing a
mono-layer of ethylene on a clean Ir(111) surface.  The sample was then heated
to 1670~K for 30~s, which resulted in large ($> 200$~nm) virtually defect-free
graphene islands. We calibrated the irradiation current with a sample bias of
40~V to reduce distortion due to secondary electrons (100~nA was used for all
samples). Taking the beam profile and sample size into account, this equals to
$0.2 − 1.0$ impacts per nm$^2$ per minute. Despite this uncertainty, arising
from the unknown exact beam profile, the relative dose should vary very little between
different samples. The irradiation was carried out perpendicular to the
graphene sheet.

\begin{figure}
\includegraphics[width=\linewidth]{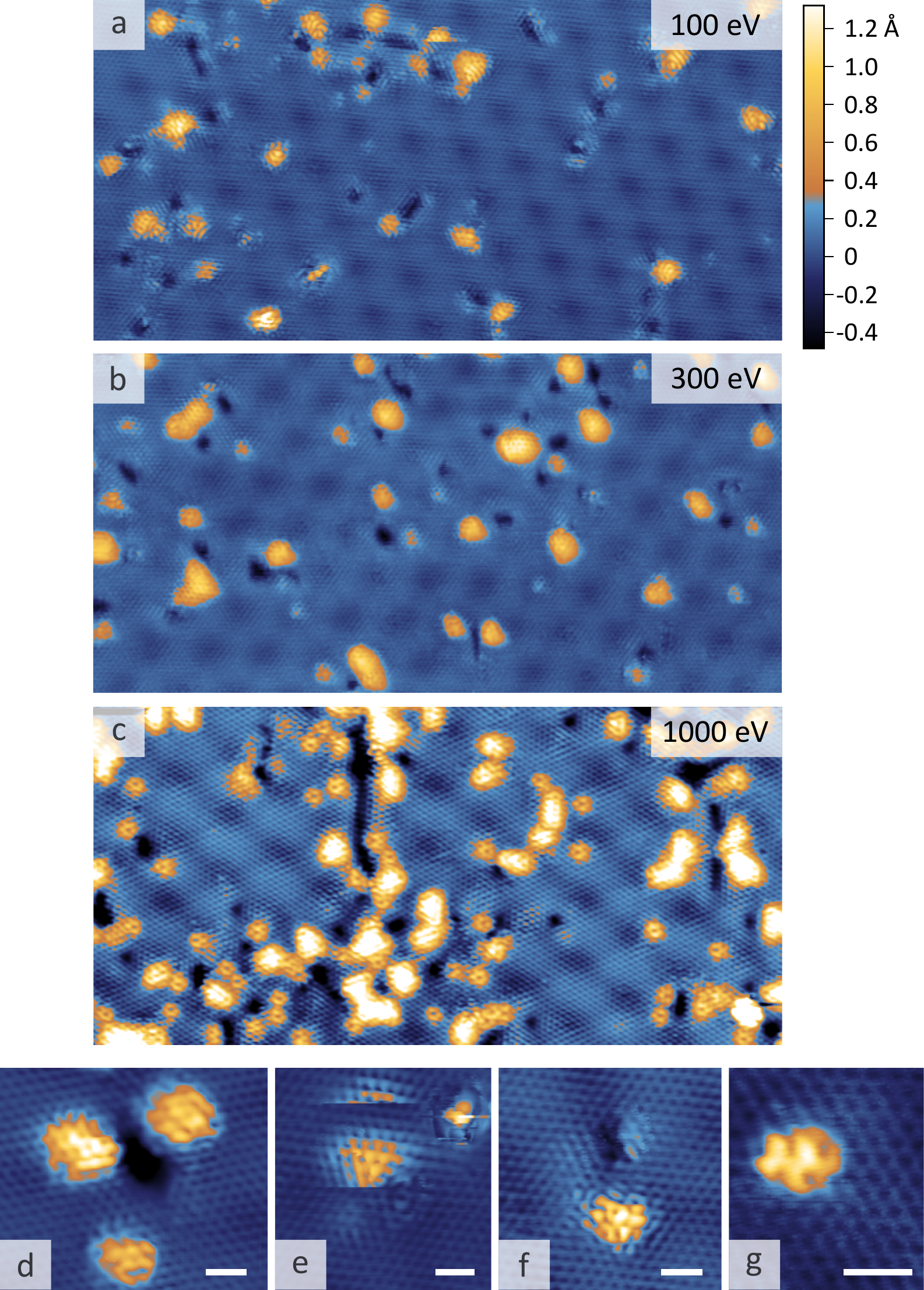}

		\caption{(Color online) STM images of the irradiated samples at different energies.
(a) 0.1~keV and (b) 0.3~keV samples were irradiated for 60~s,
(c) 1.0~keV sample for 30~s. Images are $30\times 14$~nm. STM parameters for
the images were (a) $-140$~mV/1~nA, (b) 125~mV/5~nA and (c) 125~mV/0.5~nA. (d)
Typical depression with adjacent protrusions on both sides (300~eV,
30~mV/380~pA). (e) Mobile protrusion with intact atomic lattice on top (0.1~keV,
$-140$~mV/1~nA). (f,g) Typical point-like defects (0.1~keV, $-140$~mV/1~nA and
1~keV, 260~mV/280~pA, respectively). All images were recorded at 5~K. Scale
bars in panels d-g are 1~nm.}

\label{fig::stm_overview}
\end{figure}

Atomic-resolution images showing the introduced disorder at different
experimentally studied energies (0.1, 0.3, and 1.0~keV) are presented in
Fig.~\ref{fig::stm_overview}a-c with higher magnification example images of selected
prominent features (Fig.~\ref{fig::stm_overview}d-g). Irradiation time was one
minute at 0.1~keV and 0.3~keV, and half a minute at 1.0~keV. Two kinds of
features appear in the images: darker depressed areas and adjacent bright
protrusions.  We initially interpret the depressions as defects where carbon
atoms have bound to the substrate and the protrusions as metal adatoms
intercalated under the graphene layer.  Similar interpretation was recently
made by Standop et al.~\cite{standop_ion_2013}.

% SIMULATIONS

Next, we turned to molecular dynamics (MD) to verify our interpretation of the
observed features. We followed the same approach as in our previous studies for
carbon nanotubes~\cite{tolvanen_relative_2007} and graphene (both
suspended~\cite{lehtinen_effects_2010,lehtinen_cutting_2011,ahlgren_ion_2012}
and supported~\cite{standop_ion_2013}). In our model, we replaced the iridium
substrate with platinum (as in Ref.~\cite{standop_ion_2013}), because of
the similarity of platinum and iridium in mass, structure and chemistry, and
because well-established interaction models exist for Pt-Pt and
Pt-C~\cite{albe_modeling_2002}, unlike for Ir-Ir and Ir-C. The substrate
contained 45,900 atoms set below a graphene structure (2,584 atoms) at a
distance of 3.31~{\AA}. For C-C, we used the Brenner
potential~\cite{brenner_second-generation_2002}, whereas the Ar$^+$
interactions were described with a universal repulsive
potential~\cite{ziegler_stopping_1985} (no actual charge state was considered).
Approximately 300 impact points were randomly selected for each energy. We
considered only energies $\leq 30$~keV to ensure that nuclear scattering
dominates as the damage mechanism~\cite{lehtinen_cutting_2011}. The simulations
were carried out at 0~K for 5~ps, which was enough to prevent further bonding
changes close to the interface area (energy was dissipated at the edges using
the Berendsen thermostat~\cite{berendsen_molecular_1984}). After each event, we
relaxed the atomic structure before the analysis.  To check the effect of
possible spurious structures, we conducted a further annealing study (800~K,
1~ns) for the 1~keV cases. The results did not significantly change, except for
occasional splitting of larger defects into smaller ones with approximately the
same total area and increasing bonding between the defects and the substrate.
We also extended our earlier simulations for suspended
graphene~\cite{lehtinen_effects_2010} for a direct comparison with the present
study.

\begin{figure*}
	\includegraphics[width=0.95\linewidth]{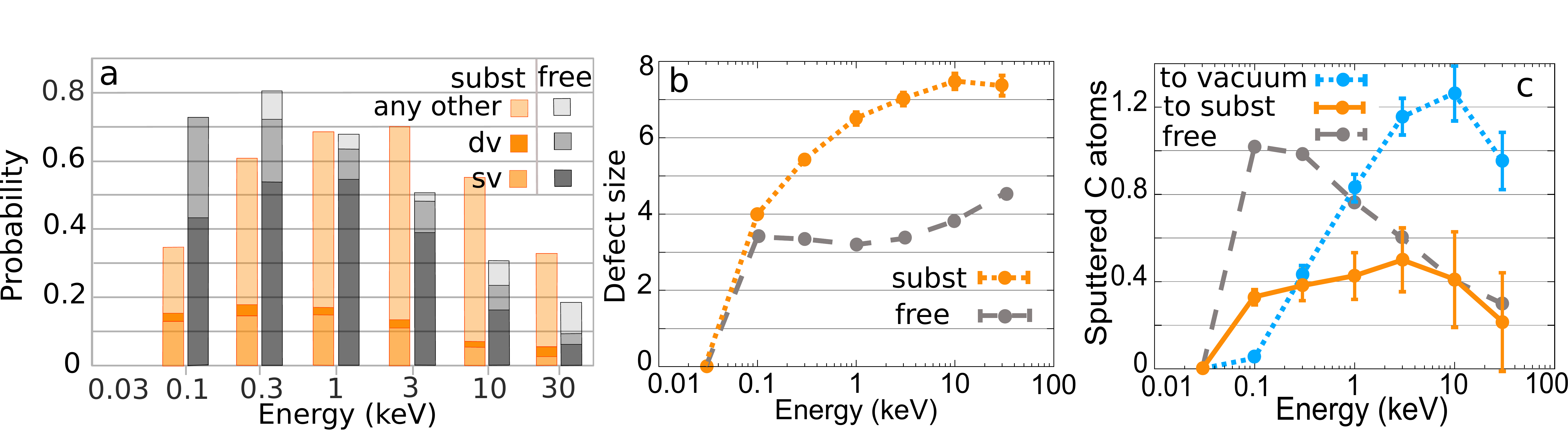}
	\caption{(Color online) Defect creation in graphene upon ion irradiation. (a)
Probabilities for creating a single vacancy (sv), double vacancy (dv) or
any other defect due to Ar$^+$ impact for supported (subst) and suspended
(free) graphene, (b) average defect size as a number of lost six-membered
carbon rings (when a defect is created), and (c) sputtering yield for carbon
atoms for suspended graphene and both towards the substrate (to
subst) and away from the substrate (to vacuum) for supported graphene as
functions of ion energy. The error bars mark the standard deviations of the data
(contained within the markers when not visible).}
 \label{fig::simdef}
\end{figure*}

We start the discussion of simulation results from the analysis of disorder created
in graphene due to the impacts (Fig.~\ref{fig::simdef}a). We categorized the
damage into single vacancies (sv), double vacancies (dv), and other defects
(identified by lost six-membered carbon rings in graphene).  The probability to
create a defect increases from zero below 100~eV up to a maximum at about
1~keV, after which it decreases because of the decreasing scattering cross
section at high energies. At the lowest energies (above 30~eV), a significant
proportion of the created defects are sv with a small probability for dv, in
agreement with experimental observations of these defects in graphene
irradiated with 140~eV Ar$^+$ on Pt(111) and SiC(0001)
surfaces~\cite{ugeda_point_2011,ugeda_electronic_2012}. At increasing energies
the share of sv's decreases while more complex defects gain prominence.

Comparing the results between supported and suspended graphene, it is clear
that the substrate leads to more complex defect structures. However, at lower
energies, it decreases the overall damaging probability. This happens because
the substrate stops the displaced low-energy carbon atoms before they are
completely detached from the graphene sheet. This trend becomes reversed at
higher energies (around 1~keV), because of the energy deposited on the top
layers of the substrate, which leads to displacement of metal atoms. These
atoms then contribute to the defect production in graphene. At the highest
studied energy (30~keV) the difference between supported and suspended graphene
starts to fade, since the bulk of the energy of the impinging ions is deposited
deep into the substrate.

The average defect size in graphene is shown in Fig.~\ref{fig::simdef}b as a
function of the ion energy.  As could be expected, we see an increase in the
defect size with increasing ion energy, but only for supported graphene, where
both sputtering from the substrate and binding to it contribute to the defect
creation. For suspended graphene, the values first rise to about three
hexagonal carbon rings (which is the size of a sv; one missing atom) per defect
at the energy of 0.1~keV, and then remain almost constant. This leveling out
occurs because the ion impact can only displace one or two carbon atoms, which
typically scatter away from the graphene plane upon further collisions, except
at very high energies (tens of keV).

We further analyzed the sputtering from graphene
(Fig.~\ref{fig::simdef}c) both towards the substrate and to the surrounding
vacuum. This data follows the same trend as the damaging probability, however
peaking at close to 10~keV. In contrast, 
the maximum sputtering yield for suspended graphene is reached at a significantly
lower energy (close to 100~eV). The reason for this is the same as previously
described: at low energies the substrate stops the displaced carbon atoms,
resulting in a lower sputtering yield. However, at high energies metal atoms
displaced from the
substrate can also detach atoms from graphene. Interestingly, after the initial
onset at 0.1~keV, the number of carbon atoms sputtering to the substrate
remains relatively constant at about 0.4 atoms/ion throughout the studied
energy range. This region coincides with the plateau in the average defect size
in the case of suspended graphene (Fig.~\ref{fig::simdef}b), because both
result mainly from carbon atoms displaced by the primary or secondary
collisions during the initial ion impact in the direction of the substrate.

\begin{figure*}
	\includegraphics[width=0.95\linewidth]{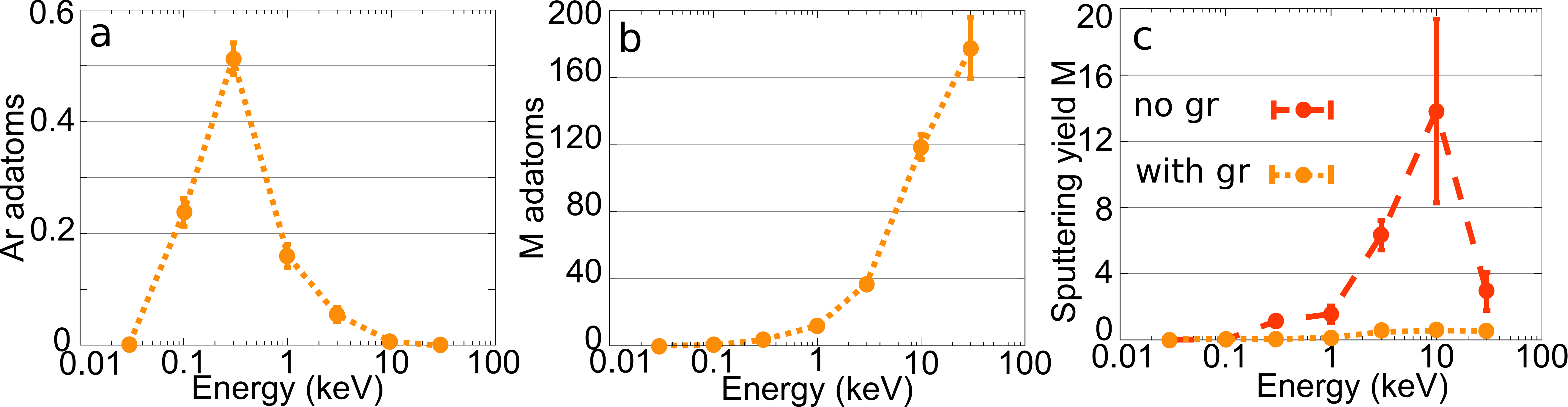}
	\caption{(Color online) Analysis of the location of Ar$^+$ ions and metal
atoms after the impact. (a) Probability for trapping the ion at the interface
area, (b) average number of substrate atoms above the metal surface per impact,
and (c) sputtering yield for the metal atoms for both a naked substrate (no gr)
and a substrate covered with graphene (with gr) as a function of ion energy.
The error bars mark the standard deviations of the data (contained within the
markers when not visible).}
	\label{fig::simsput}
\end{figure*}

Next, we turn to look at what happens to the Ar$^+$ ions after the impact. We
limit our analysis to the interface region, as only
those ions which remain at the very top of the substrate or intercalated under
graphene can be experimentally detected at the relatively low ion doses used in
this study. We classify any ion located at the top of the substrate or within
0.5~nm above it as an adatom (note that the Ar$^+$-C and Ar$^+$-Pt interactions
in our simulations are purely repulsive). Average number of these adatoms per
impinging ion is shown as a function of ion energy in
Fig.~\ref{fig::simsput}a. At 30~eV (the lowest studied energy), the ion lacks
the energy required to penetrate graphene, and is therefore always reflected
back to the vacuum. However, at 0.1~keV, already more than 20\% of the ions
become trapped at the interface. The probability increases up to about 50\% at
0.3~keV before decreasing again due to deeper penetration into the substrate.
At experimental timescales (beyond our computational reach), the Ar adatoms can
also escape through larger openings in the graphene sheet, more of which are
created at higher energies (see Fig.~\ref{fig::simdef}b).

We also counted the substrate atoms which were elevated above the surface
due to ion impacts (all atoms within 0.5~nm above the top of the
substrate were included). Average number per impact is shown as a function of
ion energy in Fig.~\ref{fig::simsput}b.  The increase is linear up to 10~keV,
after which the rate of increase decreases (note the logarithmic $x$-axis in
the plot). Obviously, several of those metal atoms trapped between graphene and
the substrate would have sputtered into vacuum, were there no graphene on top
of the metal. To get a better understanding on the role of graphene in suspending
metal sputtering, we further calculated the sputtering yield of both
graphene-covered and naked substrates as a function of the ion energy
(Fig.~\ref{fig::simsput}c). Clearly, graphene stops almost all sputtering from
the substrate, similar to what was found in the case of grazing angle
irradiation~\cite{standop_ion_2013}. Only above 1~keV some of the sputtered
metal atoms penetrate the graphene sheet, resulting in a slight deviation from
zero in the graph. This is in strong contrast with sputtering yields above
10~atoms/ion observed at intermediate energies for the naked metal substrate.

% COMPARISON

\section{Discussion}

To allow for a direct comparison between the simulations and the experiments,
we estimated the areal coverage of both the protrusions and the suppressions
from the STM images (Fig.~\ref{fig::stm_overview}) at all three energies. At
0.1~keV, the latter cover approximately 1\% of the area,
corresponding to about $0.01 − 0.05$~nm$^2$ per impact. The coverage is twice
as high (2\%) at 0.3~keV. At 1.0~keV, the coverage (with half the dose) is
about 3\%, corresponding to about $0.06 − 0.30$~nm$^2$ per impact. The
corresponding simulation results are in a very good agreement with these
estimates, being 0.036~nm$^2$, 0.086~nm$^2$ and 0.117~nm$^2$, respectively
(assuming they correspond to disorder in graphene, see
Fig.~\ref{fig::simdef}a,b). The coverage of protrusions, as estimated from the
STM images, are 6.0\%, 7.6\% and 12\% at 0.1~keV, 0.3~keV and 1.0~keV,
respectively. Corresponding relative ratios, as compared to the 1~keV case, and
taking into account the dose for each case, are 0.25, 0.32 and 1.00. These can
be compared to the number of Ar and metal atoms in the interface area in the
simulations (see Fig.~\ref{fig::simsput}a,b), which yield ratios of 0.09, 0.36
and 1.00. The discrepancy at 0.1~keV can be at least partially due to point
defects~\cite{ugeda_point_2011,ugeda_electronic_2012} that appear as
protrusions in STM images (see also Fig.~\ref{fig::stm_overview}f,g) and/or Ar
adatoms, which are trapped at the interface and only observed at 0.1~keV in
the experiments (see Fig.~\ref{fig::stm_overview}d and Supplementary
Material presented below). Simulations indicate some Ar adatoms also at 0.3~keV
and 1.0~keV, which we expect to be due to the short time scale, which does not
allow the atoms to escape through larger holes in graphene created at higher
energies. Both of these effects would increase the apparent ratio of
protrusions at 0.1~keV, as compared to higher energies.

\begin{figure*}
\includegraphics[width=0.95\linewidth]{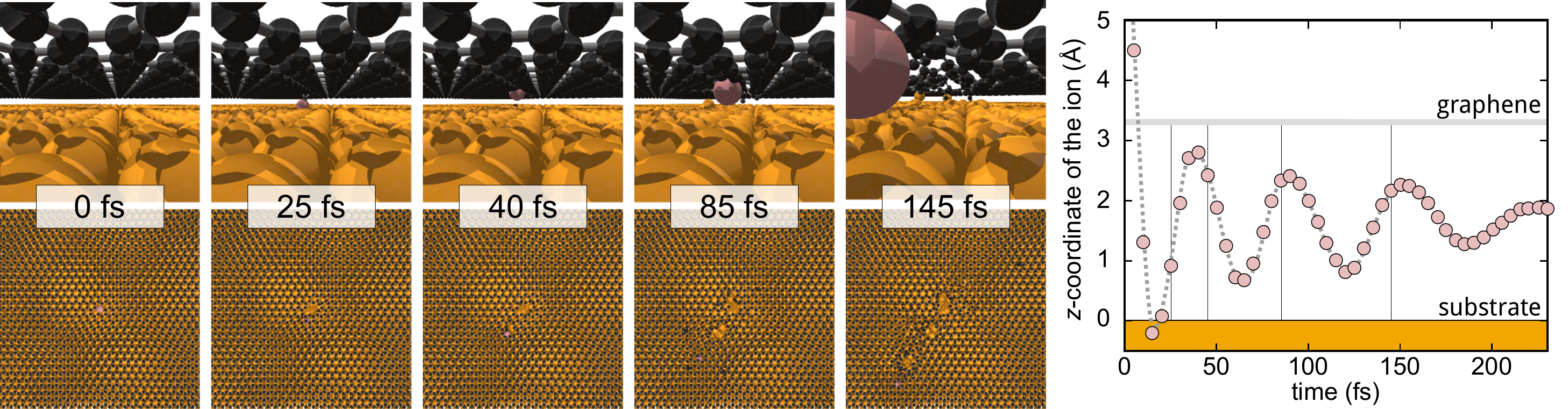}

\caption{
(Color online) Creation of a line-defect due to an Ar$^+$ impact at 1~keV.
After the initial penetration of the graphene sheet, the ion is deflected to
the space between the graphene and the metal substrate, and bounces from one to
the other while constantly producing damage along the way (upper row:
perspective view, lower row: top view). In the plot, the $z$-coordinate of the
ion is shown as a function of simulation time. The vertical lines correspond to
the times of the frames presented in the snapshots.  }

\label{fig::linedef}
\end{figure*}

Finally, we also point out the long line-like depressions, which appear in the
STM images after 1~keV irradiation (for an example, see
Fig.~\ref{fig::stm_overview}c). These defects are remarkably similar to the ion
tracks caused by grazing-angle irradiation~\cite{standop_ion_2013}, which can
be turned into a graphene nanomesh via high-temperature annealing. Although
counter-intuitive for perpendicular irradiation, as in our case, it turns out
that the formation mechanism is exactly the same in both cases: the incoming
ion is occasionally deflected into the interface area between the substrate and
graphene, and causes a line of small vacancy-type defects to form while
channeling between the two materials (see Fig.~\ref{fig::linedef} and
Supplementary Material).

% CONCLUSIONS

\section{Conclusions}

As a conclusion, we have carried out a study of Ar$^+$ irradiation of graphene on
a metal substrate combining atomistic simulations and scanning-probe
microscopy. Our results show that the presence of a substrate leads to more
complicated defect structures as compared to suspended graphene, and that the
complexity of the created defects can be controlled via irradiation energy: on
the one hand, at 1~keV, even irradiation perpendicular to the graphene sheet
leads to formation of line-like defects via interface channeling of the
impinging ion; on the other hand, lower energies result in simpler defects,
such as single vacancies in graphene (up to one third of the defects at
100~eV). Our results provide a basis for controlled introduction of disorder
into graphene on a metal substrate, and may open the way towards graphene-metal
interface structures with tailored properties.

\section*{Acknowledgments}

We acknowledge Austrian Science Fund (FWF): M~1481-N20, Helsinki University
Funds, the European Research Council (ERC-2011-StG No.~278698 "PRECISE-NANO"),
the Finnish Academy of Science and Letters, the Academy of Finland (Centre of
Excellence in Low Temperature Quantum Phenomena and Devices No. 250280) and the
Finnish Cultural Foundation for funding. Further, we acknowledge CSC Finland
and Vienna Scientific Cluster for generous grants of computational resources.
We also thank Arkady Krasheninnikov and Carsten Busse for insightful
discussions.

% BIBLIOGRAPHY

\newpage

\begin{appendix}

\section{Supplementary Material}

This supplement contains additional experimental images and spectra of the Ar
adatoms observed under the graphene sheet after 100~eV irradiation.

\begin{figure}[hb!]

\includegraphics[width=0.98\linewidth]{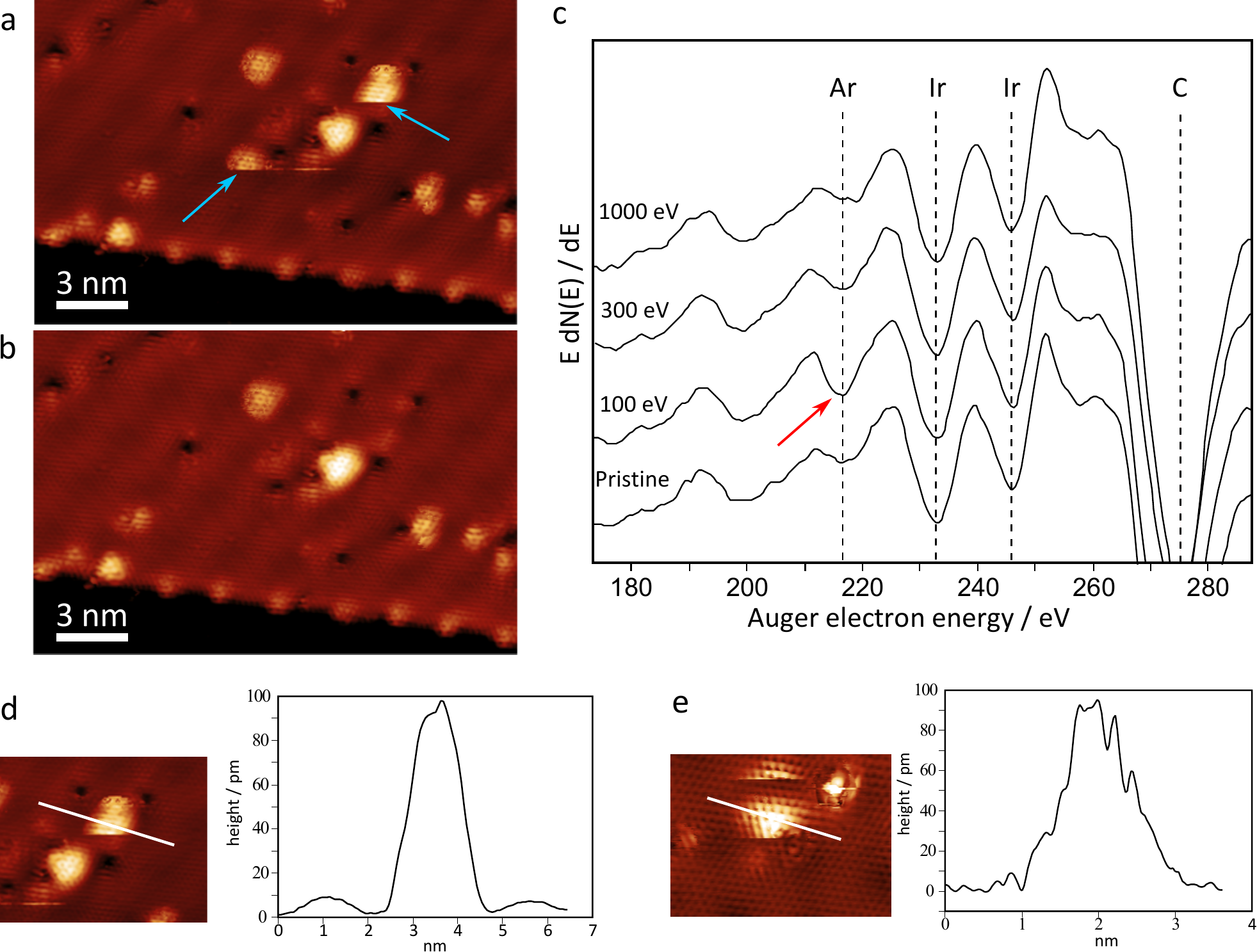}

\caption{(a,b) Two consecutive STM scans (-70~mV / 1~nA) showing moving
defects, which disappear during the scans. (c) Auger electron spectrum of
pristine graphene on Ir and samples irradiated with different energies. The
100~eV data shows a clear dip at the Ar LMM Auger transition. The smaller
feature close to this one arises from the Ir spectrum.  (d,e) Height profiles
over two different mobile protrusion showing an apparent height of 1~{\AA} for
both.}

\label{sf::ar_adat}
\end{figure}

Figs.~\ref{sf::ar_adat}a and \ref{sf::ar_adat}b show two consecutive STM scans
(scan direction down) where two protrusions (marked by blue arrows) disappear
from under the STM tip while scanning. Unlike in Fig.~1a of the main article,
where the protrusion moved back and forth while scanning, here the protrusions
disappear completely. Our interpretation is that the intercalated Ar adatoms
are pushed out through holes in the graphene sheet. The apparent STM height of
all mobile protrusions is about 1~{\AA} (see Fig.~\ref{sf::ar_adat}~d,e).  

To provide further evidence that the mobile protrusions we see are actually
intercalated Ar atoms, we measured Auger electron spectra (AES) of samples
irradiated with different energies with the same absolute dose
(Fig.~\ref{sf::ar_adat}c). The spectra were acquired in situ with a Perkin
Elmer 15-255G double pass cylindrical mirror analyzer (DPCMA) using a 3~keV
electron beam. The dip in the derivative spectrum at the Ar LMM transition is
clearly the largest in the 100~eV sample and diminishes almost completely with
increasing energy. The AES signal of the Ar LMM transition comes from the few
top atomic layers, as the effective attenuation length of the Auger electrons
in iridium is only 4~{\AA}~\cite{powell_nist_2011}. Hence the AES data shows
that as the ion energy increases, Ar is deposited deeper into the sample.  To
ensure a measurable Ar signal, the AES experiments were conducted with a
slightly higher dose, compared to the STM experiments.

\end{appendix}

\end{document}